\documentclass[11pt]{article} 

\usepackage[utf8]{inputenc} 
\usepackage{amsmath, amssymb, graphics}
\usepackage{url}

\newcommand{\mathsym}[1]{{}}
\newcommand{\unicode}[1]{{}}
\newcommand{\tab}{\hspace*{3em}}

\usepackage{geometry} 
\usepackage{graphicx} 

\usepackage{booktabs} 
\usepackage{array} 
\usepackage{paralist} 
\usepackage{verbatim} 
\usepackage{subfig} 

\usepackage{fancyhdr} 
\usepackage{sectsty}
\allsectionsfont{\sffamily\mdseries\upshape} 

\usepackage[nottoc,notlof,notlot]{tocbibind} 
\usepackage[titles,subfigure]{tocloft} 

\title{Optical Geometry of the Kerr Space-time}
\author{Cameron Bloomer}
\date{May 15th, 2011} 

\begin{document}
\maketitle

\vspace{25 mm}

\abstract{}

We pursue a geometrical approach to gravitational lensing theory. We present a survey of the background theory of General Relativity, including particular properties of the Schwarzschild and Kerr solutions. Next we outline a proof of the Gauss Bonnet theorem and its applications to surfaces in optical geometry, as developed by G. W. Gibbons and M. C. Werner.  Finally, we attempt to extend this geometrical approach to the axially symmetric Kerr spacetime, and arrive at an expression for the gravitational deflection angle in the equatorial plane.

\pagebreak
\section{Introduction}

\tab In 1916, Einstein published ``The Foundation of the General Theory of Relativity," an extension of his special theory of relativity to include gravity. From a contemporary perspective, physicists salute this achievement as the inauguration of twentieth-century physics. Indeed, entire theories of scientific advancement as paradigm shifts may take the 1916 paper as their model: so Kuhn's notion of paradigm shift seems to stem from this impulse.\cite{kuhn} Thus, the development of physics and the history of science derive from Einstein's radically new and intuitively surprising claims. In addition, the paper impelled new directions in experimentation.\cite{howard} Indeed, it was not long until a series of experiments verified his most radical thoughts. The theory came in a series of papers from 1911 to 1916, and was followed by the empirical research.\\
\tab Einstein's theory of gravity successfully reproduced Newtonian gravitation in domains with relatively weak gravitational fields. Moreover, it explained the extra 43 arcseconds per century in the precession of Mercury's orbit, previously unaccounted for. One of the most striking predictions of Einstein's theory of General Relativity is the deflection of light due to the curvature of spacetime. In 1919, Eddington's observation and confirmation of this effect made Einstein an instant international celebrity.\\
\tab While Newton's theory of gravity established the idea of force, instead Einstein viewed mass itself as equivalent to the curvature of spacetime, in a precise sense. In contrast to all other forces, no known object is immune to the effects of gravity. Paths previously consider to be curved by Newton -say the orbit of the moon- were described by Einstein simply as straight lines in a curved spacetime, or geodesics. Therefore, while Newton held that light, composed as he believed of massless particles, would travel in a straight line through empty space, Einstein's theory implied the path of light, indeed of all objects, should curve around a massive body. This is the gravitational lensing effect. In 1919 Sir Arthur Eddington, then secretary of the Royal Society and chief assistant to the Astronomer Royal of the Greenwhich observatory, received papers from Willem de Sitter describing Einstein's theory. He organized two expeditions to test the gravitational lensing effect by   observing a solar eclipse. Despite Eddington's unfounded assumption that the data fall into one of three values, the expedition was considered the second piece of experimental support for Einstein's remarkable theory, following the calculation of the precession of Mercury in 1915.\cite{efstathiou} \\
\tab  The first known intergalactic lens, ``Twin QSO," was discovered in 1979.\cite{wambsganss} Since then hundreds of microlensing events have been identified, spurring research and even allowing the detection of previously invisible celestial objects. Gravitational lensing theory studies geometrical optics in spacetime, connecting astrophysics, theoretical physics, and mathematics. Recent research develops a new method of analyzing lensing systems, one which connects the differential geometry of these systems with their topology.\cite{werner194} This paper will attempt to extend this work from spherically symmetric spacetimes to the Kerr spacetime. \\

\pagebreak
\section{The Schwarzchild spacetime}

\tab In 1916, shortly after Einstein published his theory of G.R., Karl Schwarzschild found an exact solution to Einstein's equations that describes the unique structure of a nonrotating spherically symmetric spacetime. The Schwarzschild metric successfully reproduces Newtonian gravity in the non-relativistic (weak field) limit. Moreover, it allowed a calculation of the precession of the orbit of Mercury, accounting for the extra 43 arcseconds per century that had long been observed. The Schwarzschild metric also allowed the slight deflection of light by the sun to be calculated, and this was soon followed by Sir Arthur's expedition in 1919, which observed such deflection during a solar eclipse. This provided a strong confirmation of Einstein's theory of General Relativity, published just three years before. The components of the  metric $g$ are given in terms of the line element, $ds^{2} = g_{\mu \nu} dx^{\mu} dx^{nu} $, as follows. \\

\large
${ds}^{2}  =  -\left(1-\frac{2 M}{r} \right) dt^2 + \frac{1}{c^2}\left(1-\frac{2 M}{r}\right)^{-1}dr^2+ \frac{r^2}{c^2} \sin ^2(\theta)d\phi^2+\frac{r^2}{c^2}{d\theta}^2$\\
\normalsize

We choose units to set the speed of light $c$ = 1, and we will follow the convention of using Latin indices to denote space coordinates, and Greek indices to denote spacetime coordinates. Since light is massless, light rays will follow null geodesics in accordance with G.R. Since the Schwarzschild metric is spherically symmetric, we may restrict ourselves to the equatorial plane without loss of generality, letting $\theta = \frac{\pi}{2}$  and \text{d$\theta$} = 0. Fermat's principle states that light rays follow a path of critical time. Furthermore, a spatial projection of the null geodesics will produce the light rays.\cite{werner194} We find that the following metric describes the optical reference geometry: \\

\large
\tab $ dt^{2}= \frac{r^{2}}{{(r - 2 M)}^{2}} dr^2 + \frac{r^3}{(r - 2 M)} d\phi^{2}$\\

\tab $g^{opt}_{mn}=g_{mn} \left/ \left(-g_{00}\right)\right.$\\
\normalsize

\noindent Although our optical surface has only (2) spatial components, let us define in general a \large $\Gamma _{\mu \nu }^{\lambda }$,\\

\large
\tab$\Gamma _{\mu \nu }^{\lambda } \text{ := } \frac{1}{2} {g}^{\lambda \rho }\left(\partial _{x^{\nu }}g_{\rho \mu } + \partial
_{x^{\mu }}g_{\rho \nu } - \partial _{x^{\rho }}g_{\mu \nu }\right)$\\
\normalsize
\\
Where a sum is implied over $\rho$, in keeping with the Einstein summation convention, which we shall continue to use. This gamma essentially represents the $\lambda$ - component of the covariant derivative of the derivative of the $\mu$ basis vector with respective to the $\nu$ basis vector. To be precise:\\

\large
\tab $\Gamma _{\mu \nu }^{\lambda } \partial_{\lambda} = \nabla_{\partial_{\mu}} \partial_{\nu}$\\
\normalsize

\noindent Next we calculate the Gauss Curvature $K$ of this optical surface.

\subsection{Calculation of the Gauss Curvature}

\tab The Riemann curvature tensor $R$ measures the failure of a vector to return to its original position following parallel transport about some infinitesimal loop in a given plane. The tensor can be defined on any manifold $\mathcal{M}$ with an affine connection, and on a two dimensional manifold this tensor can be shown to have only one independent component.\cite{oprea} Although the Gauss Curvature $K$ can be defined as the product of the two principal curvatures $k_1$ and $k_2$, which are the eigenvalues of the second fundamental form of $\mathcal{M}$, Gauss's Theorema Egregium proves that $K$ is in fact an intrinsic measure of curvature, and $K$ can be expressed purely in terms of the metric. For a surface, it is not hard to show that the independent component of the Riemann tensor, say $R_{1212}$, is related to the Gauss Curvature $K$ by $K = R_{1212}/g$, where $g$ is the determinant of the metric.\cite{oprea} We wrote the following Mathematica code to calculate the Gauss Curvature of the optical surface of the Schwarzschild spacetime.\\

\noindent \tab$\text{Clear}[X,r,t, \mu , \Gamma ,\text{RAbcd},\text{Rabcd}];$\\
\tab$X\text{  }= \{r, \phi \};$\\
\\
\tab $g = \{\{r{}^{\wedge}2 / (r - 2 M){}^{\wedge}2,0 \},\{0,r{}^{\wedge}3/(r- 2 M)\}\};$\\
\tab $gi = Inverse[g];$\\
\\
\tab $\Gamma [\lambda \_, \mu \_, \nu \_] \text{:=} \text{Sum}\left[1/2 \text{gi}[[\lambda ,\rho ]]\left(\partial _{X[[\nu ]]}g[[\rho ,\mu ]] + \partial
_{X[[\mu ]]}g[[\rho ,\nu ]] - \partial _{X[[\rho ]]}g[[\mu ,\nu ]]\right), \{\rho ,2\}\right];$\\
\\
\tab $\text{RAbcd}[\text{a\_}, \text{b\_}, \text{c\_}, \text{d\_}] \text{:=}\partial _{X[[c]]}\Gamma [a,d,b] - \partial _{X[[d]]}\Gamma [a,c,b] +\\ \\
\tab \tab \tab \tab \text{Sum}[\Gamma [a,c,\lambda ] \Gamma [\lambda ,d,b],\{\lambda ,2\}] - \text{Sum}[\Gamma [a,d,\lambda ] \Gamma [\lambda ,c,b],\{\lambda ,2\}];$\\
\\
\tab $\text{Rabcd}[\text{a\_},\text{b\_},\text{c\_},\text{d\_}] \text{:=} \text{Sum}[g[[a,z]]\text{RAbcd}[z,b,c,d], \{z,2\}];$\\
\\
\tab $K[\text{a\_},\text{b\_},\text{c\_},\text{d\_}]\text{:=} \text{Rabcd}[a,b,c,d]/(g[[a,c]]g[[d,b]]-g[[a,d]]g[[c,b]]);$\\
\\
\tab $\text{FullSimplify}[K[1,2,1,2]]$\\

\noindent We find that the Gauss Curvature is:\\

\large
\noindent \tab $K = \frac{M (3 M-2 r)}{r^4}$\\
\normalsize

\noindent This expression varies only with the radius $r$, and the mass-like constant $M$, as we expect.

\subsection{The geodesic equations}

\tab Geodesic curvature is an intrinsic measure of the amount a curve is curving within a manifold. In General Relativity, all objects, in the absence of forces, follow a path of zero geodesic curvature (geodesics). Define a curve $C:\mathbb{R} \rightarrow M$, such that $p \mapsto x\left(p\right)$. The geodesic is then defined to be the non-normal component of the change in the curve $C$'s direction and speed: \\ 

\large
\tab $\kappa _g \text{:=} <\nabla _{x'}x',x''/|x''|> = g\left(\nabla _{x'}x',x''/|x''|\right)$\\
\normalsize

\noindent From this definition and the one for $\Gamma_{\nu \lambda }{}^{\mu }$, one arrives at the following differential equation, which defines a geodesic curve $x$, parameterized by the affine parameter $p_{1}$.\\
\\
\large
\tab $\frac{dx^{\mu }}{dp_{1}^{2}} + \Gamma _{\nu \lambda }^{\mu }\frac{x^{\nu }}{dp_{1}} \frac{x^{\lambda }}{dp_{1}} = 0$\\
\normalsize
\\
We define the metric coefficients $E$, $F$, and $G$ as follows:  \\
\\
\tab $E \text{:=} <\text{dr}\text{, }\text{dr}>  =  \frac{r^2}{(r - 2 M)^2}$\\
\tab $F\text{:=} <\text{dr} \text{, }\text{d$\phi $}>  =  0$\\
\tab $G\text{:=} <\text{d$\phi $} \text{, }\text{d$\phi $}>  = \text{  }\frac{r^3}{(r - 2 M)}$\\
\\
Clearly, $E_{\phi } = G_{\phi } = 0$. This means our patch is Clairaut in r and so the geodesic equations for light, in the optical geometric surface of the Schwarzschild equatorial plane, reduce to:\\
\\
\tab $r'' + \frac{\partial _rE}{2 E}(r')^2-\frac{\partial _rG}{2 E}(\phi ')^2= 0$\\
\tab $\phi'' + \frac{G_{\phi }}{G}r'\phi ' = 0$\\
\\
\tab These differential equations do have an exact solution; however, there is a different method of deriving the asymptotic deflection and of light passing by the Schwarzschild black hole, without having to solve or integrate the geodesic equations of motion. This method is due to \cite{werner194}, and manages to avoid having to deal with the curvature singularity of the black hole altogether.

\pagebreak
\section{Geometrical approach to lensing}

\tab Now we will apply the method developed in \cite{werner194} to find the asymptotic deflection angle of a light ray, with impact parameter b, that passes by a Schwarzschild black hole of mass $m = \frac{G M}{c^2}$. We will utilize the Gauss-Bonnet theorem, which elegantly connects a surface's geometry to its topology. The Gauss-Bonnet theorem then states that given a 2-dimensional manifold $\mathcal{M}$, with piecewise smooth boundary $\partial \mathcal{M}$, interior angles $i_j$, geodesic curvature $\kappa_g$, Gauss curvature $K$, and Euler Chi function $\chi \left(\mathcal{M}\right)$: \\

\large
$$\int\int _{\mathcal{M}} K dA  + \int _{\partial{ \mathcal{M}}}  \kappa _g  ds  +  \sum _j {(\pi - i_j)}= 2\pi  \chi (\mathcal{M})$$
\normalsize
\\

Following \cite{werner194}, we define the gravitational lensing system in the following way, and seek to find the asymptotic deflection angle of a light beam:\\

\includegraphics[height=60mm]{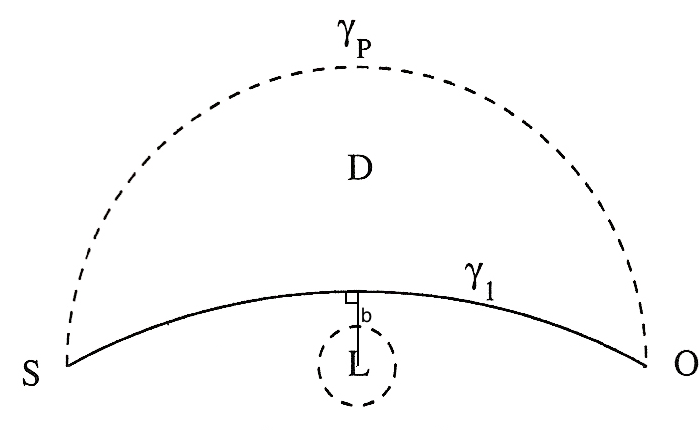} \\
Figure 1.\\
\\
\tab Let S be the source of the light, O be the position of the observer, and $L$ be the spherically symmetric mass. Let $\gamma_1$ be a light geodesic from S to O, and let $b$ be the impact parameter of $\gamma_1$ with respect to L.  Let $\gamma_p$ be part of a sector of a circle centered at L and intersecting $\gamma_1$ at S and O at right angles. Let $\phi$ be the angular coordinate and let $r$ be the radial coordinate, both centered on L. Let $\delta$ be the asymptotic deflection angle of $\gamma_1$. Let $r_S$ and $r_O$ be the radial coordinates of S and O, respectively. We shall assume that $b$, $r_S$, and $r_O$ $\gg$ $m$, so that the geodesic $\gamma_1$ is always in the weak field limit. Furthermore, we assume $r_S$, $r_O$ $\gg$ $b$, so that $\gamma_1$ may be approximated as the union of two straight segments.\cite{werner194} \\
\tab Now let us find $\delta$. Let $D$ be the region bounded by $\gamma_1$ and $\gamma_p$. Let the boundary $\partial D$ be $\gamma_1$ traversed from S to O and $\gamma_p$ from O to S. The angular range of $\gamma_1$ is $\pi + \delta$ and $\chi(D) = 1$. We apply the Gauss-Bonnet theorem to find:  \\

\large
\noindent \tab $\int\int _{D}K dA + \int _0^{\pi  + \delta }\kappa _g\left(\gamma _p\right) \left(\frac{dt}{d\phi} \right) d\phi + 2 \frac{\pi}{2} = 2 \pi$\\
\normalsize
\\
\large
\noindent \tab $\int _0^{\pi  + \delta }\kappa _g\left(\gamma _p\right) \left(\frac{dt}{d\phi} \right) d\phi  - \pi  = -\int\int _{D}K dA.$\\
\normalsize

We are interested in the asymptotic deflection angle, and so since the Schwarzschild metric is asymptotically flat, over $\gamma_p$ we can take $\kappa_g \left( \gamma_p \right) \frac{dt}{d\phi} = 1$. Similarly, we can describe $\gamma_1$ as $r\left(\phi\right) = b / sin(\phi$).  \cite[p. 3]{werner194} The equation then becomes: \\

\large
\noindent \tab $\delta = -\int\int_{D} K dA$\\ \\
\tab $dA = \sqrt{det\left(g^{opt}\right)} dr d\phi = \sqrt{\frac{r^{5}}{{\left( r - 2 M\right)}^{3}}} dr d\phi$\\ \\
\tab $\delta = -\int\int_{D} \frac{M (3 M-2 r)}{r^4}  \sqrt{\frac{r^{5}}{{\left( r - 2 M\right)}^{3}}}dr d\phi$\\ \\
\tab $	\approx	\int_{0}^{\pi}\int_{b/sin\left(\phi\right)}^{\infty}  \frac{2 M}{r} \frac{\left(1 - \frac{3 M}{2 r}\right)}{\left(1 - \frac{2 M}{r}\right)}dr d\phi$\\ \\
\tab $\approx \frac{4 M}{b}. $\\
\normalsize

This matches the known formula for the deflection angle. See \cite{werner194} for a fuller exposition of this geometrical approach to gravitational lensing. We will next examine the Gauss-Bonnet theorem in more detail, providing the main steps of a proof.

\pagebreak
\section{The Gauss-Bonnet theorem}

\tab Let $\mathcal{M}$ be a 2-dimensional manifold with boundary $\partial{\mathcal{M}}$, and with interior angles $i_j$. Then the Gauss-Bonnet theorem relates the integral of the Gauss curvature, plus the total angle curved out by the boundary of M to the Euler Chi function in the following way: 

\large
$$\int _{\partial{\mathcal{M}}}  \kappa _g  ds  +  \int\int _{\mathcal{M}} K dA  + \sum _j \left(\pi  - i_j\right)= 2\pi  \chi (\mathcal{M})$$
\normalsize

\noindent{To prove this, first let $\alpha : [0,1] \ \rightarrow \partial{M}$ be a simple closed unit-speed curve, parameterized by $0 \leq  s \leq  2 \pi $. Let $\mathcal{M}$ be parameterized by $x\left(r,\phi\right)$, with $X(u,v) \in T_x \mathcal{M}$. Let \\}

\noindent{\tab $X_u  := \partial _u X$\\ 
\tab $X_v := \partial _v X$\\

\noindent{We can assume the parameterization has been chosen so that $g\left(X_u, X_v\right) =0$. Let\\}

\noindent{\tab $E := g\left(X_u, X_u\right)$\\ \\
\tab$G :=g\left(X_v, X_v\right)$\\ \\
\tab$E_1 := \frac{X_u}{\sqrt{\left|g\left(X_u, X_u\right)\right|}}$\\ \\
\tab$E_2 := \frac{X_v}{\sqrt{\left|g\left(X_v, X_v\right)\right|}}$\\ \\

\noindent{By construction, we see that $g\left(E_1,E_2\right) = 0$, and $g\left(E_1,E_1\right)=g\left(E_2,E_2\right)=1$. So we have an orthonormal basis. Define:\\}

$\omega _{\text{ij}} = \nabla _{\alpha '}E_{j }E_i $\\

$\omega _{21} = \nabla _{\alpha'} E_1  E_2$\\

$\cos\left(\theta\right) := g\left( \alpha ',E_1\right)$\\

$\alpha' = \cos\left(\theta\right) E_1 + \sin\left(\theta\right) E_2$\\

$\nabla _{\alpha '}\alpha ' = \left((d\theta )/ (dt)+ \omega _{21}\right)\left(-\sin\left(\theta\right) E_1 + \cos\left(\theta\right)E_2\right)$\\

$\kappa _g := g\left(\nabla _{\alpha '}\alpha ', \alpha '\right) = \left(\frac{d\theta}{dt}+ \omega _{21}\right)\left(\sin\left(\theta\right)^{2} + \cos\left(\theta\right)^{2}\right) = \frac{d\theta} {dt}+ \omega _{21}$\\

$\alpha ' = X_{u }u' + X_v v'$\\

$\nabla _{\alpha '}E_{1 }\text{ } E_2 = \left[\left(E_1\right)_u du/ds + \left(E_1\right)_v dv/ds\right]  E_2$\\

$=\left(\left[X_{uu}/\sqrt{E} - X_u \left.E_u\right/(2 E^{3/2})\right] \frac{du}{ds} + \left[X_{uv}/\sqrt{E} - X_u \left.E_v\right/(2
E^{3/2})\right] \frac{dv}{ds}\right)  \frac{X_v} {\sqrt{G}}$\\ \\
\noindent{Note:\\ \\}
\tab $X_{uu}\text{ }X_v = -\frac{E_v}{2}$\\
\tab $X_{uv} \text{ }X_v = \frac{G_u} {2}$\\ \\}
\noindent{Therefore\\ \\
\tab $\omega _{21}= -E_v/\left(2 \sqrt{E G}\right) \frac{du}{ds} + G_u/\left(2 \sqrt{E G}\right) \frac{dv}{ds}$\\}

\noindent{Recall the formula for the Gauss curvature K in terms of the metric:\\}

\large
\quad \quad \quad $K = -\frac{1}{2\sqrt{E G}} \left(\left(\frac{E_v}{\sqrt{E G}}\right)_v + \left(\frac{G_u}{\sqrt{E G}}\right)_u\right)$\\ \\
\normalsize 
Now,\\ \\
\tab$\omega _{21} = \kappa _g - \frac{d\theta}{ds}$\\ \\
\tab$\frac{d\theta}{ds} - \kappa _g = E_v/\left(2 \sqrt{E G}\right) \frac{du}{ds} - G_u/\left(2 \sqrt{E G}\right) \frac{dv}{ds}$\\ \\
\tab $\int \frac{d\theta}{ds} ds - \int \kappa _g ds = \int  \left(E_v/\left(2 \sqrt{E G}\right) \frac{du}{ds} - G_u/\left(2 \sqrt{E G}\right) \\
\frac{dv}{ds}\right)ds $\\

\noindent{\tab$=\int _{\alpha }E_v/\left(2 \sqrt{E G}\right) du - G_u/\left(2 \sqrt{E G}\right) dv$\\ \\ }

\noindent{Now, applying Green's theorem, we get:\\ \\}

\noindent{\tab$= \int \int _{\mathcal{M}}-\frac{1}{2} \left[\left(\frac{G_u}{\sqrt{E G}}\right)_u + \left(\frac{E_v}{\sqrt{E G}}\right)_v \right]du dv.$}\\ \\

\noindent{\tab$ = \int \int _{\mathcal{M}}K \sqrt{E G}du dv$\\ \\}
\noindent{\tab$ = \int\int _{\mathcal{M}}KdA$\\}

\noindent{So, \\ \\}
\noindent{\tab $\int_{\partial \mathcal{M}} \kappa _g ds+\int\int _{\mathcal{M}}KdA = \int \frac{d\theta} {ds} \, ds = \theta \left(2\pi \right) - \theta \left(0\right)$\\ \\}

Furthermore, if we allow a piecewise-continuous boundary to our manifold, then we have corners which contribute to the total angle subtended, giving: \\ \\
$\int_{\partial \mathcal{M}} \kappa _g ds+\int\int _{\mathcal{M}}KdA + \sum _j \left(\pi  - i_j\right) =\theta \left(2\pi \right) - \theta (0) = 2\pi$ an integer.\\ \\

We shall argue that the integer multiplying $2 \pi$ above, call it  $\chi(\mathcal{M})$, is indeed the Euler Chi function. Suppose we have some triangulation of $\mathcal{M}$. Let $T$ be the set of all triangles. We can sum the equation above for every triangle $t \in T$. Note that each triangle is a topological disk. We therefore find that the sum of these equations is equal to $2 \pi F$, where F is the number of triangles.\\ \\

$\sum_{t} \left[\int_{\partial t} \kappa _g ds+\int _{t}KdA + \sum _j \left(\pi  - i_{tj}\right)\right] = 2\pi F$\\ \\

Note that the sum of the integrals of $K$ over the triangles' areas is equal to the integral of $K$ over the whole area of the manifold. Now note every interior edge of the triangulation is traversed once by a triangle in one direction, and once by the bordering triangle in the other direction. Therefore the two integrals of $\kappa_g$ over such an interior edge cancel each other, and so their sum is zero; furthermore, if we remove such an interior edge, we have reduced the number of faces by one. Therefore we can change the sum of the integral of $\kappa_g$ over the boundaries of the triangles, to an integral of $\kappa_g$ over the boundary of the manifold, provided we also subtract $2 \pi$ times the number of interior edges from the right side of the equation. Now note that every interior vertex contributes a total of $2 \pi$ to the sum of the interior angles of the triangles. Therefore let us remove these interior vertices from the sum of the interior angles, and simply add $2 \pi$ times the number of interior vertices to the right hand side of the equation. To complete the proof, we note that the number of exterior edges must be equal to the number of exterior vertices, since the boundary of the manifold is a union of topological circles. Hence $\chi(\mathcal{M}) = V - E + F$, as desired. Therefore, we have found:\\

\large
\tab $\int _{\partial{\mathcal{M}}} \kappa _g ds  +  \int\int _{\mathcal{M}} K dA  + \sum _j \left(\pi  - i_j\right)=  2\pi  \chi (\mathcal{M}). \tab Q.E.D.$\\ \\
\normalsize
Proof based from \cite[p. 281-291]{oprea}.\\

\section{The Kerr Metric}

\tab The Kerr metric is an exact solution of Einstein's equations which describes the axially symmetric spacetime of a rotating spherically symmetric mass. It is a more realistic model of the spacetime of a black hole, as real black hole generally have non-zero angular momentum.\\
\tab 	An accurate formula for the asymptotic deflection angle of light around the Kerr black hole could provide a second direct method of measuring the angular momentum of black holes, and could thereby be used to test the idea of cosmic censorship, proposed by physicist Dr. Roger Penrose.\cite{werner2}\\
\tab Cosmic censorship is the idea that the universe does not allow ``naked" singularities. Singularities are the Achilles heal of Einstein's theory of General Relativity since the theory breaks down there--the Riemann curvature tensor diverges. Physicists Dr. Penrose and Dr. Hawking have proposed that such singularities must be enclosed in a (black) event horizon to keep them from causally interacting with the rest of the universe. If this were not the case, eventually one would no longer be able to say what will happen anywhere in the universe after a certain time. \cite{werner194}\\
\tab The Kerr metric is as follows:\\
	
$ds^{2}= -\frac{ \Delta -a^2 \sin^{2}\left(\theta\right)}{\rho ^{2}}dt^{2}-\frac{4 M a r \sin^{2}\left(\theta\right)}{\rho^{2}}dt d\phi + \\\tab \frac{\left(a^{2}+r^{2}\right)^{2}-a^{2} \Delta  \sin^{2}\left(\theta \right)}{\rho^{2}}  \sin^{2}\left(\theta \right) d\phi^{2} + \frac{ \rho^{2}}{\Delta }dr^{2}+ \rho^{2}d\theta^{2} $      \cite[p. 309]{schutz} \\

\noindent $\Delta := a^{2}-2 M r+r^{2}$\\
\noindent $\rho^{2} := r^{2} + a^{2} \cos^2\left(\theta\right)$\\

\noindent{For the sake of simplicity, we shall restrict ourselves to the equatorial plane, so that $\theta = \frac{\pi}{2}$ and $d\theta = 0$.  Because the metric is reflection-symmetric about $\theta = \frac{\pi}{2}$, a trajectory which begins with $p^{\theta} =0$ will continue to have $p^{\theta} = 0.$\cite[p. 359]{schutz} Therefore a geodesic that begins in the equatorial plane will remain in it. The metric g becomes: \\}

 \noindent{\tab $ds^{2} := g_{\mu \nu }\text{dx}^{\mu }\text{dx}^{\nu }$\\}

\noindent{\tab$= -\frac{ \left(\Delta -a^{2}\right)}{r^{2}}dt^{2}+\frac{ \left(\left(a^{2}+r^{2}\right)^{2}-a^{2} \Delta \right)}{r^{2}}d\phi^{2}-\frac{4 a M}{r}dt d\phi+\frac{r^{2}}{\Delta }dr^{2}$\\}

\noindent We again wish to find null geodesics, so we set $ds^2 = 0$, and solve for $dt$ in order to find an optical metric. \\

\noindent{\tab$\frac{ \left(\Delta -a^{2}\right)}{r^{2}}dt^{2} + \frac{4 a M}{r}dt d\phi = \frac{r^{2}}{\Delta}dr^{2}$\\}

\noindent{\tab$\left(\sqrt{r^{2}-2 M r}dt +\frac{2 a M r}{\sqrt{r^{2}-2 M r}}d\phi\right)^{2}=\frac{r^{4}}{\Delta }dr^{2}+ \left(r^{4}
+a^{2} r^{2}+2 a^{2} M r+ \frac{4 a^{2} M^{2} r}{(r-2 M)}\right)d\phi^{2}$\\}

\noindent{\tab $dt = \sqrt{\frac{r^{3}}{\Delta  (r-2 M)}dr^{2}+\frac{r^{2} \Delta }{(r-2 M)^{2}}d\phi^{2}}-\frac{2 M a}{(r-2 M)}d\phi$.\\}

\noindent{This asymmetrical form of metric was first discovered and studied by Randers.\cite{randers} We have found our optical Finsler metric for the optical reference geometry of the Kerr. Further, \cite{bao} gives several formulas for the Finsler equivalent of Gauss curvature, as well as a computational algorithm for computing it. \\}

\subsection{Calculation of Gauss Curvature} 

\tab We hypothesize that the leading-order term in $a = J^2 / M$, of the optical deflection angle of the Kerr may come not from the asymmetrical Finsler part of the metric, but rather from the pseudo-Riemannian part. The following Mathematica code was written to arrive at an estimate for the Kerr deflection angle based on the pseudo-Riemannian (diagonal) part of the metric, ignoring the Finsler contribution, in order to test this hypothesis.\\ 

\noindent \tab$\text{Clear}[X,r,t, \mu , \Gamma ,\text{RAbcd},\text{Rabcd}];$\\
\noindent \tab$X = \{r,\phi \};$\\ \\
\tab$\Delta  = r^{2} - 2 M r + a^{2};$\\
\tab$\text{$\rho^2$} = r^2;$\\ \\
\tab$g = \{\{r^{3} / (\Delta  (r - 2 M)),0\},\{0,(r^{2} a^{2} + r^{3}(r - 2 M))/ (r - 2 M)^{2}\}\};$\\  \\
\tab$\text{gi} = \text{Inverse}[g];$\\
\\
\tab$\Gamma [\lambda \_, \mu \_, \nu \_] \text{:=} \text{Sum}\left[1/2 \text{gi}[[\lambda ,\rho ]]\left(\partial _{X[[\nu ]]}g[[\rho ,\mu ]] + \partial
_{X[[\mu ]]}g[[\rho ,\nu ]] - \partial _{X[[\rho ]]}g[[\mu ,\nu ]]\right), \{\rho ,2\}\right];$\\
\\
\tab $\text{RAbcd}[\text{a\_}, \text{b\_}, \text{c\_}, \text{d\_}] \text{:=}\partial _{X[[c]]}\Gamma [a,d,b] - \partial _{X[[d]]}\Gamma [a,c,b] +\\ \\
\tab \tab \tab \tab \text{Sum}[\Gamma [a,c,\lambda ] \Gamma [\lambda ,d,b],\{\lambda ,2\}] - \text{Sum}[\Gamma [a,d,\lambda ] \Gamma [\lambda ,c,b],\{\lambda ,2\}];$\\
\\
\tab$\text{Rabcd}[\text{a$\_$},\text{b$\_$},\text{c$\_$},\text{d$\_$}] \text{:=} \text{Sum}[g[[a,z]]\text{RAbcd}[z,b,c,d], \{z,2\}];$\\
\\
\tab$K[\text{a$\_$},\text{b$\_$},\text{c$\_$},\text{d$\_$}]\text{:=} \text{Rabcd}[a,b,c,d]/(g[[a,c]]g[[d,b]]-g[[a,d]]g[[c,b]]);$\\
\\
\tab$\text{FullSimplify}[K[1,2,1,2]]$\\ 

\noindent We arrive at the following expression for the Gauss Curvature $K$ of our reduced metric:\\

\large
\noindent{\hspace{2em}$K = \frac{M \left(6 a^{2} \left(-M+r\right)+r \left(6 M^{2}-7 M r+2 r^{2}\right)\right)}{\left(2 M-r\right) r^{5}}$}\\
\normalsize
\\

\noindent We find that the limit as $a \rightarrow 0$ of this $K$ reduces to the value found for the Gauss K of the optical geometry of Schwarzschild, as expected.\\

\subsection{Extension of the Geometric Approach} 

\tab Now we attempt to extend the geometric approach which we applied to the Schwarzschild spacetime to the Kerr spacetime. 
\noindent{Starting from Gauss-Bonnet and using the expression for $K$ above, we find:\\}

\noindent{\tab $\int _0^{\pi + \delta}\kappa _g\left(\gamma _p\right)\frac{dt}{d\phi} d\phi  - \pi  = -\int\int _{D}K dA$}}\\ 

\noindent{$\tab \delta  =-\int\int _{D}K dA$}\\ 

\noindent{\tab $dA = \sqrt{\left| g_{opt} \right|}dr d\phi = \sqrt{\left(\frac{r^{3}}{\Delta \left(r-2M\right)}\right)\left(\frac{r^{2} \Delta} {\left(r-2M\right)^{2}}\right)}dr d\phi $}\\ 

\noindent{$\tab \delta  =-\int\int _{D}K  \frac{r^{5/2} }{(r-2M)^{3/2}}drd\phi$\\

\noindent \tab $\text{ }\approx -\int _0^{\pi }\int _{b/\text{Sin}[\phi ]}^{\infty }K \frac{r^{5/2} }{(r-2M)^{3/2}}drd\phi $\\ \\ 
\tab$= \int _0^{\pi }\int _{b/\text{Sin}[\phi ]}^{\infty }\frac{M r^{5/2}(6a^{2}(r-M) + 2 r^3 (1- 2M/r)(1 - 3 M/2r))}{r^{15/2}(1-2M/r)^{5/2}}dr d\phi$\\ \\
\noindent{\tab$= \int _{0}^{\pi }\int _{b/\text{Sin}[\phi ]}^{\infty }M 6 a^{2} (1-M/r)/(r^{4} (1-2 M/r)^{5/2})dr d\phi \text{   }+ $}\\ 
\\ 
\tab$\int _0^{\pi }\int _{b/\text{Sin}[\phi ]}^{\infty }2 M (1-2M/r)(1-3M/2r)/(r^2 (1-2M/r)^{5/2}) dr d\phi $\\
\\ 
\noindent \tab $\approx  \int _0^{\pi }\int _{b/\text{Sin}[\phi ]}^{\infty }6a^2 M r^{-4}dr d\phi  + \int _0^{\pi }\int _{b/\text{Sin}[\phi
]}^{\infty }2 M r^{-2}dr d\phi $\\
\\ 
\noindent{\tab$= \int _0^{\pi }\int _{b/\text{Sin}[\phi ]}^{\infty }6a^2 M r{-4} dr d\phi  + 4M/b$}\\
\\
\noindent{\tab$= 2a^2 M /b^3 \int_0^{\pi } \text{Sin}[\phi ]^3 \, d\phi  + 4 M/b$}\\
\\
\tab$=2M a^2/b^3  (4/3) + 4M/b$\\
\\
\tab$= \frac{4M}{b}\left(1+\frac{2}{3}\frac{a^2}{b^2}\right)\text{     }\square  $\\
\\
And so we have found the contribution to the deflection angle from the pseudo-Riemannian part of the metric shows up at second order in $a.$ \\

\section{Discussion}
\tab In \cite{aazami} and\cite{sereno}, expressions are derived for the deflection angle of light curving around a Kerr blackhole, in the weak-deflection limit, to third order in $a$ and in $M$, where $M$ is the scaled mass of the blackhole, and $a=\frac{J^2}{M}$ is a measure of the angular momentum. Let $b$ be the impact parameter. Restricting their expression to the equatorial plane, the angle of deflection for a retrograde photon reduces to:\\

$\delta$ =  $\frac{4M}{b} + (\frac{15}{4} \pi M^{2} + 4 M a) /b^{2} + (\frac{128}{3} M^{3} + 10 \pi M^{2} a + 2 M a^{2}) / b^{3}$.\cite[p. 5]{sereno}\\
 
We note the second order term in $a$, $2 \frac{M}{b^{3}} {{a}}^{2}$, is reasonably close to the answer we got from the Riemannian part of the metric: $\frac{8}{3} \frac{M}{b^3} {{a}}^{2}$; however, the first order terms in $a$ do not come from the Riemannian part of the metric, but rather from the Finsler geometry. Indeed, while the diagonal part of the metric does not distinguish between a positive displacement in $\phi$ and a negative one, in reality we should find that retrograde light rays, moving against the rotation of the blackhole, should have an increased deflection angle, while prograde light rays should have a decreased deflection angle. This is caused by the light ray spending more time near the blackhole in the retrograde case, and less time close to the black hole in the prograde case.\cite{sereno} If we define $S^{+}_{-}$ to equal $+1$ for a retrograde ray, and $-1$ for a prograde one, the expression for the magnitude of the asymptotic deflection angle from \cite{sereno} becomes:\\

\noindent \tab $\|\delta\| = \left[\frac{4 M}{b} + \frac{15 \pi}{4} \frac{M^2}{b^2} + \frac{128}{3} \frac{M^3}{b^3}\right] + S^{+}_{-}\left[ \frac{4 M}{b^2} + \frac{10 \pi M^2}{b^3}\right] a +  \left[ \frac {2 M }{b^3} \right] a^2 $\\

So we see that retrograde light rays have a larger deflection angle than prograde ones, and that while the terms linear in $a$ depend on which side of the blackhole the light ray passes, the $a^2$ term has no such dependence. 

\pagebreak{}	
\section{Conclusion}

\tab The curvature of the Kerr and Schwarzschild spacetimes is negative. This should imply that locally parallel geodesics diverge. The topology of these spacetimes is therefore essential to  the convergence of light rays, and hence to the gravitational lensing effect. The Gauss-Bonnet theorem, which relates the local geometry of a two-dimensional manifold to its global topological structure, directly connects the global properties of the optical geometry, necessary for gravitational lensing, to the measurable deflection angles. The theorem thereby provides an effective, informative framework for examining lensing.  \\ 
	\tab In conclusion, we have found that the Gauss curvature derived from the pseudo-Riemannian part of the Kerr metric does not account for the first-order effect of the angular momentum quantity a = $\frac{J^{2}}{M}$, but instead appears at second order in $a$. Perhaps future work could examine whether the Gaussian curvature analogue in Finsler geometry can be used in a manner similar to the Gauss curvature, to deduce a measure of the asymptotic optical deflection angle in the Kerr spacetime.

\section{Acknowledgments}
\tab I would like to thank Dr. Marcus Werner for his all help.
\pagebreak


\begin{thebibliography}{9}
\bibitem{aazami}
A. B. Aazami, C. R. Keeton, A. O. Petters.
``Lensing by Kerr Black Holes. II: Analytical Study of Quasi-Equatorial Lensing Observables."
\emph{arXiv}: 1102.4304v1 [astro-ph.CO] (2011).
\bibitem{bao}
D. Bao, S. S. Chern, Z. Shen. 
\emph{An Introduction to Reimann-Finsler Geometry.} Graduate Texts in Mathematics 200. Springer-Verlag, New York. 2000.
\bibitem{boyer}
R. H. Boyer, R. W. Linquist.
``Maximal Analytic Extension of the Kerr Metric."
\emph{Journal of Mathematical Physics}, Volume 8, Number 2. 1967:265-281.
\bibitem{efstathiou}
G. Efstathiou.
Lecture notes from ``Theory of Relativity." Cambridge University, Michaelmas Term 2002:89-147.
\bibitem{howard}
Don Howard.
``The Development of Twentieth-Century Philosophy of Science." In Michael Janssen and Chistoph Lehner, eds, \emph{Cambridge Companion to Einstein}. Cambridge. 2004:1-44.
\bibitem{kuhn}
Thomas S. Kuhn. 
\emph{The Structure of Scientific Revolutions.} 1st. ed. Chicago: Univ. of Chicago Press. 1962.
\bibitem{oprea}
J. Oprea.
\emph{Differential Geometry and Its Applications.} Second Edition.
The Mathematical Association of America, USA. 2007.
\bibitem{randers}
G. Randers.
``On an Asymmetrical Metric in the Four-Space of General Relativity."
 \emph{Physical Review}, Volume 59. 1941:195-199.
 \bibitem{schutz}
 B. Schutz.
 \emph{A First Course in General Relativity.} Second Edition.
 Cambridge University Press, Cambridge. 2009.
 \bibitem{sereno}
M. Sereno, F. De Luca.
``Analytical Kerr Black Hole Lensing in the Weak Deflection Limit."
\emph{Physical Review,} D 74, 123099 (2006).
\bibitem{wambsganss}
Joachim Wambsganss.
``Gravitational Lensing in Astronomy."
\emph{Living Review in Relativity}. Potsdam, Germany. 2001.
\bibitem{werner194}
M. C. Werner, G. W. Gibbons.
``Applications of the Gauss-Bonnet Theorem to Gravitational Lensing."
 \emph{Classical Quantum Gravity}, 25, 235009 (2008).
 \bibitem{werner2}
 M. C. Werner, A. O. Petters.
``Magnification Relations for Kerr Lensing and Testing Cosmic Censorship."
\emph{Physical Review}, D 76, 064024 (2007).
\end{thebibliography}
\end{document}